\documentclass[12pt]{article}
\usepackage{epsf}
\usepackage{amsmath}

\usepackage{graphics}
\usepackage{cite}

\begin{titlepage}

\begin{document}

\hfill February 2022

\begin{center}

{\bf \LARGE Entropy of the Universe and Hierarchical Dark Matter}\\
\vspace{2.5cm}
{\bf Paul H. Frampton}\footnote{paul.h.frampton@gmail.com}\\
\vspace{0.5cm}
{\it Dipartimento di Matematica e Fisica "Ennio De Giorgi",\\ 
Universit\`{a} del Salento and INFN-Lecce,\\ Via Arnesano, 73100 Lecce, Italy.
}

\vspace{1.0in}

\begin{abstract}
\noindent
We discuss the relationship between dark matter and the entropy of the universe
with the premise that dark matter exists in the form of primordial black holes (PBHs) in a
hierarchy of mass tiers. The lightest tier are intermediate-mass PIMBHs within galaxies including the
Milky Way. Supermassive black holes at galactic centres are in the second tier. We are led 
to speculate that there exists a third tier of
extremely massive PBHs, more massive than entire galaxies. 
We discuss future observations by the Rubin Observatory and the James
Webb Space Telescope.

\end{abstract}

\end{center}
\end{titlepage}

\section{Introduction} 
\bigskip

\noindent
In particle theory, the concept of entropy is not regarded as fundamental. Particle 
theorists who have published hundreds of papers may never
need the word entropy. For one elementary particle entropy
is neither defined nor useful.

\bigskip

\noindent
In general relativity and cosmology, the situation is different. For black holes,
entropy is a central and useful concept. For cosmology, the entropy of the universe
has often been considered, although not often enough in our opinion. We shall
argue that the origin and nature of cosmological dark matter can be best understood
by consideration of the entropy of the universe. We have made such an
argument four years ago\cite{Frampton2018} but that discussion was perhaps too diluted
by considering simultaneously dark matter being made from elementary particles such
as WIMPs and axions, as were favoured three decades ago\cite{KolbTurner}.

\bigskip

\noindent
In this paper, we dispose of microscopic candidates in one paragraph.
The standard model of particle theory (SM) has two examples of lack of naturalness, the
Higgs boson and the strong CP problem. Our position is that to understand these
we still need to understand better the SM itself. Regarding the strong
CP problem, it is too {\it ad hoc} to posit a spontaneously broken global
symmetry and consequences which include an axion. Concerning the
WIMP, the idea that dark matter experiences weak interactions arose
from assuming TeV-scale supersymmetry which is now disfavoured
by LHC data. To identify the dark matter, we instead look up.

\bigskip

\noindent
Assuming dark matter is astrophysical, and that the reason for its existence lies
in the Second Law of Thermodynamics, we shall be led uniquely to
the dark matter constituent as the Primordial Black Hole (PBH). We must admit
that there is no observational evidence for any PBH, but according to our discussion
PBHs must exist. In the ensuing discussion, we shall speculate that they exist
in abundance in three tiers of mass up to and including at several galactic masses.

\bigskip

\noindent
Because PBH entropy goes like mass squared, we are mainly interested in masses satisfying
$M_{PBH} >100 M_{\odot}$. Within the Milky Way, we use the acronym PIMBH for intermediate
mass PBHs in the mass range $10^2 M_{\odot} < M_{PIMBH} < 10^5 M_{\odot}$. Outside the
Milky Way we entertain all masses $10^2 M_{\odot} < M_{PBH} < 10^{17} M_{\odot}$. Of these,
we use PSMBH for supermassive PBHs in the mass range $10^5 M_{\odot} < M_{PSMBH} < 10^{11} M_{\odot}$
and PEMBH for extra massive PBHs with $10^{11}M_{\odot} < M_{PEMBH} < 10^{17} M_{\odot}$.

\bigskip

\noindent
Although the visible universe (VU) is not a black hole, its Schwartzschild radius is about
68\% of its physical radius, 30 Gly versus 44 Gly, so it is close. This curious fact seems to
have no bearing on the nature of dark matter.
A few more acronyms will be useful: CMB. CIB and CXB. CMB is the familiar cosmic microwave
background while I and X refer to Infra-red and X-ray respectively.

\bigskip

\noindent
The arrangement of the paper is as follows. We discuss entropy and the second law in
sections 2 and 3 respectively, then in section 4 primordial black holes. In sections 5 and 6
we discuss two methods of PBH detection, microlensing and cosmic infrared background
respectively. Finally in section 7 is a discussion.

\section{Entropy}

\noindent
We begin with the premise that the early universe be regarded in an approximate sense
as a thermodynamically-isolated system for the purposes of our discussion.
It certainly contains a number of particles, $\sim 10^{80}$, vastly larger than the
numbers normally appearing in statistical mechanics, such as Avogadro's number,
$\sim 6 \times 10^{23}$ molecules per mole.

\bigskip

\noindent
No heat ever enters or leaves and it can be considered as though its surface were
covered by a perfect thermal insulator. It is impracticable to solve all the Boltzmann
transport equations so it is mandatory to use thermodynamic arguments, provided
that we may argue that the system is proximate to thermal equilibrium.

\bigskip

\noindent
Making the then-unsupported assumption in 1872 \cite{Boltzmann} of atoms and
molecules, Boltzmann discovered the quantity $S(t)$ in terms of the 
molecular momentum distribution function $f(\textbf{p},t)$
\begin{equation}
S(t) = -  \int  d \textbf{p} f(\textbf{p}, t) \log f(\textbf{p}.t)
\end{equation}
which satisfies
\begin{equation}
\left( \frac{d S(t)}{dt} \right) \geq 0
\label{SLT}
\end{equation}
and can be identified with the thermodynamic entropy.
The crucial inequality, Eq(\ref{SLT}), the Second Law, was derived in
\cite{Boltzmann} for non-equilibrium systems assuming only the 
Boltzmann transport equations and the ergodic hypothesis.

\bigskip

\noindent
Ascertaining the nature of the dark matter can be regarded as a detective's mission
and there are useful clues in the visible universe. In \cite{Frampton2018}, we 
made an inventory of the entropies of the known objects in the visible universe,
using a venerable source, the book \cite{Weinberg1972}. Let us model the visible
universe as containing $10^{11}$ galaxies each of mass $10^{12} M_{\odot}$ and
each containing one central SMBH with mass $10^7 M_{\odot}$. We recall the
dimensionless entropy of a black hole $S/k (M_{BH}=\eta M_{\odot}) \sim 10^{78} \eta^2$.
Then the inventory is
\begin{itemize}
\item SMBHs $\sim 10^{103}$
\item Photons $\sim 10^{88}$
\item Neutrinos $\sim 10^{88}$
\item Baryons $\sim 10^{80}$
\end{itemize}

\bigskip

\noindent
We regard this entropy inventory as a {\bf first clue}. From the point of view of entropy t
he universe would be only
infinitesimally changed if everything except the SMBHs were removed. This suggests that
more generally black holes totally dominate the entropy, as we shall find in the sequel.

\bigskip

\noindent
A second remarkable fact about the visible universe is the near-perfect
black-body spectrum of the CMB which originated some 300,000 years
after the beginning of the present expansion era, or after the Big Bang
in a more familiar language. We are not tied to a Big Bang which could
well be replaced by a bounce in a cyclic cosmology.

\bigskip

\noindent
The precise CMB spectrum is a {\bf second clue} about dark matter. It suggests
that the plasma of electrons and protons prior to recombination is in
excellent thermal equilibrium, and hence the matter sector
was in thermal equilibrium for the first 300,000 years.
This, combined with the thermal isolation mentioned already,
underwrites the use of entropy, and the second law, during this period.

\bigskip

\noindent
A {\bf third clue} and final one about dark matter lies with the holographic principle
\cite{Hooft} which provides, as upper limit on the entropy of the visible
universe, the area of its surface in units of the Planck length.
Given it present co-moving radius $44$ Gly this requires
$S/k \leq 10^{123}$. The entropy of the contents which is so bounded
might nevertheless tend to approach\cite{CFK} a limit which is
many orders of magnitude higher
than the total entropy in the limited inventory listed above.

\section{Second Law}

\noindent
For primordial black holes (PBHs) formed at cosmic time $t$, their mass 
may be taken to be governed by the horizon size, giving
\begin{equation}
M_{PBH} = 10^5 M_{\odot} \left( \frac{t}{1 \sec} \right)
\label{Mpbh}
\end{equation}
so that PBHs with masses $10^2 M_{\odot} < M_{PBH} < 10^{17}M_{\odot}$
are produced for $10^{-3}s < t < 30ky$. The top few orders of
magnitude are unlikely, but possible.

\bigskip

\noindent
A tendency to increase the entropy of the universe towards
$S_U/k \sim 10^{123}$ can be most readily be achieved by
the formation of PBHs, the more massive the better, because
$S_{BH}/k \sim 10^{78} \eta^2$ for mass $M_{BH}= \eta M_{\odot}.$
For example, in the case that a PEMBH existed with
$M_{PEMBH} \sim 10^{17} M_{\odot}$ it would have
$S/k \sim 10^{112}$ which is a billion times the entropy of the items 
listed in our previous inventory.

\bigskip

\noindent
The PBH mass function is unknown so we must make
reasonable conjectures which may approximate Nature.
For a preliminary discussion we may take monochromatic
distributions separately for PIMBHs, PSMBHs and PEMBHs.
The real mass function is expected to be smoother
but the general features in our discussion of entropy should
remain valid.

\bigskip

\noindent
In a toy model for the visible universe we include $10^{11}$
galaxies each with mass $10^{12}M_{\odot}$. As a hierarchical
dark matter we shall take as illustration all PIMBHs with
$100 M_{\odot}$; all PSMBHs with $10^7 M_{\odot}$;
all PEMBHs at $10^{14} M_{\odot}$. Let the number of each
type be $n_I$, $n_S$ and $n_E$, respectively.
The total dark matter mass is then
\begin{equation}
M = \left( 10^2 n_I + 10^7 n_S + 10^{14} n_E \right) M_{\odot}
\label{Mass}
\end{equation}
while the total entropy contributed by all PBHs is
\begin{equation}
S/k = \left( 10^{82} n_I + 10^{92} n_S + 10^{106} n_E \right)
\label{Entropy}
\end{equation}

\bigskip

\noindent
Let us begin with the middle one of the three hierarchical
tiers, the supermassive black holes known to reside in
galactic centres. In our toy model, $n_S$ is equal to the
number of galaxies $n_S = 10^{11}$ so their total mass and 
entropy are, from Eq.(\ref{Mass}),
\begin{equation}
M(PSMBHs) =  10^{18} M_{\odot}
\label{MassPSMBHs}
\end{equation}
and, from Eq.(\ref{Entropy}),
\begin{equation}
S(PSMBHs)/k = 10^{103} 
\label{EntropyPSMBHs}
\end{equation}

\bigskip

\noindent
Before considering Eqs.(\ref{Mass}) and (\ref{Entropy}) further,
let us step back and ask which of the three terms in each
equation is most likely to be dominant? The answer is
different for Eqs.(\ref{Mass}) and (\ref{Entropy}) because
entropy $S/k$ and mass $M$ have the relationship $S/k \propto M^2$.

\bigskip

\noindent
The total mass in Eq.(\ref{Mass}) is comparable to the total
mass of the visible universe which is $ \sim 10^{123} M_{\odot}$.
Comparison with $M(PSMBHs)$ in Eq.(\ref{MassPSMBHs})
then show that the second term in the R.H.S. of Eq.(\ref{Mass}) is sub-dominant, being several orders of magnitude less than the L.H.S.

\bigskip

\noindent
Now let us discuss the first term on the R.H.S.  In our toy model
every galaxy has mass $10^{12} M_{\odot}$ which is dominated
by the dark matter halo made up of $100 M_{\odot}$ PIMBHs and
therefore, since there are $10^{11}$
galaxies, we take $n_I = (10^{11})\times(10^{10}) =10^{21}$
whereupon the total mass and 
entropy of the PIMBHs are, from Eq.(\ref{Mass}),
\begin{equation}
M(PIMBHs) =  10^{23} M_{\odot}
\label{MassPIMBHs}
\end{equation}
and, from Eq.(\ref{Entropy}),
\begin{equation}
S(PIMBHs)/k = 10^{103} 
\label{EntropyPiMBHs}
\end{equation}

\bigskip

\noindent
From Eq.(\ref{MassPIMBHs}) we deduce that the first term
on the R.H.S. of Eq.(\ref{Mass}) is a dominant term. We already
know that the second term on the R.H.S. is relatively small.
What about the third and last term? At this stage, we can
say little except that observation is consistent with it vanishing.
Perhaps surprisingly, to jump ahead, after discussion of the
entropy equation, Eq.(\ref{Entropy}), we shall suggest the
third term on the R.H.S. of Eq.(\ref{Mass}) is comparable to
the first term on the R.H.S. of Eq.(\ref{Mass}), thus
providing a rather novel viewpoint of dark matter.

\bigskip

\noindent
Substituting our choices $n_I= 10^{21} M_{\odot}$
and $n_S = 10^{11}$ into Eq.(\ref{Entropy}) we find for 
the total entropy
\begin{equation}
S/k = \left( 2 \times 10^{103} + 10^{106} n_E \right)
\label{EntropyPBHs}
\end{equation}
to be compared to the total mass
\begin{equation}
M= \left( 10^{23} + 10^{18} + 10^{14}  n_E \right) M_{\odot}
\label{MassPBHs}
\end{equation}

\bigskip

\noindent
In Eq.(\ref{MassPBHs}), for consistency we must bound the
parameter $n_E$ from above by $n_E \leq 10^9$ to avoid
overclosing the universe. It is interesting to study the upper
limit of $n_E$ in the entropy equation, Eq.(\ref{EntropyPBHs}).
This gives $\sim 10^{115}$ to be compared with the holographic bound on the
entropy\cite{Hooft,CFK} which is $\sim 10^{123}$.

\bigskip

\noindent
In the absence of any observational evidence about either
dark matter or primordial black holes, we need to look at the visible universe 
from the two theoretical viewpoints of mass
and entropy. This suggests the most likely scenario which
is $n_E \sim 10^9$. This predicts that our toy universe contains
of order one billion extra-massive black hole with masses 
$O(10^{14}M_{\odot})$ or perhaps a smaller number of even more
massive PBHs.  Because of their extraordinarily high masses,
these PEMBHs are not expected to be
associated with a specific galaxy or cluster of galaxies.

\section{Primordial Black Holes}

\noindent
If black holes make up all the dark matter, they cannot be
all gravity-collapse black holes because of baryon mumber
conservation. The amount of dark matter is more than five times
that of baryons. Therefore, most or all dark-matter black
holes must instead be primordial.

\bigskip

\noindent
PBHs are black holes formed in the early universe when there 
is s high density and sufficiently large fluctuations and inhomogeneities.
Their
existence was first conjectured in the 1960s in the Soviet
Union\cite{Novikov} and independently in the 1970s, in 
the West\cite{CarrHawking}. Initially it was realised that
only PBHs with mass greater than $10^{-18} M_{\odot}$
could survive until the present time because of Hawking
evaporation. Nevertheless, it was generally assumed
that PBHs were all very much lighter than the Sun and
hence even more lighter than all the PBHs considered in
this paper.

\bigskip

\noindent
During this early era of extremely light PBHs, the seminal
idea that PBHs could form all the dark matter was proposed
in 1975 by Chapline\cite{Chapline1975}. In 2009
\cite{FHKR2009} and in 2010 
\cite{ChaplineEntropy}
the relevance of entropy in cosmological evolution
emerged.

\bigskip

\noindent
Beginning in 2010\cite{FKTY}, the upper limit on PBH mass was removed by
showing that in a specific model of hybrid inflation, with two stages
of inflation, a parametric resonance could mathematically yield
fluctuations
and inhomogeneities of arbitrarily large size. We regard this as merely
an existence theorem and that such formation might take place without
inflation.

\bigskip

\noindent
The possibility of PBHs with many solar masses led to the 
2015 dark matter proposal
in \cite{Frampton2015} that PIMBHs provide an excellent astrophysical
candidate for dark matter in the Milky Way halo, especially given the
absence of a compelling elementary particle candidate either within
the standard model or in any plausible extension thereof. This was
further underscored in \cite{ChaplineFrampton2016}. Both of these
papers emphasised microlensing by PIMBHs of starlight from
the Magellanic Clouds \cite{MACHO} as a promising method for detection
of PIMBHs in the Milky Way.

\bigskip

\noindent
These PIMBHs are now to be regarded as the first of three mass tiers,
the second being the supermassive PSMBHs at galactic centres and the
third being extremely massive PEMBHs, more massive than galaxies.

\bigskip

\noindent
Returning to our thermodynamic arguments about entropy, we use the
entropy inventory of the known entities to observe the idea that very
massive black holes already dominate entropy through the PSMBHs
which we assume are primordial because there seems to be insufficient cosmic
time for stellar mass black holes adequately to grow by accretion and mergers.

\bigskip

\noindent
For the entropy of the universe to be nearer to its holographic
upper limit, we are led to introduce $10^9$  PEMBHs of
$10^{14} M_{\odot}$ to reach $S/k \sim 10^{115}$.  To achieve
the maximum $S/k \sim 10^{123}$ is possible with just ten
PEMBHs of $10^{22} M_{\odot}$ which, if true, would be revolutionary.

\section{Microlensing}

\noindent
Gravitational lensing of a distant star by a nearer massive object
or lens, moving across the field of view,
gives rise to an enhancement of the star and to a temporal light curve
whose duration is proportional to the square root of the mass of the lens,
as displayed in Eq.(\ref{duration}).

\bigskip

\noindent
Aa already mentioned, a direct way to discover PIMBHs in the Milky Way would be to use
microlensing\cite{Frampton2015,ChaplineFrampton2016}  of light from
the stars in the Magellanic clouds. Assuming a
transit velocity $200$km/s an estimate of the
duration $\hat{t}$ of the light curve at half
maximum is
\begin{equation}
\hat{t} \sim 0.2y \left( \frac{M_{lens}}{M_{\odot}} \right)^{\frac{1}{2}}
\label{duration}
\end{equation}
which means that for $10^2M_{\odot} < M_{PIMBH}< 10^5 M_{\odot}$
the duration of the light curve is in the range
$2y < \hat{t} < 60y$. Masses below $2,500 M_{\odot}$ with
$\hat{t}<10y$ are clearly the most practicable to measure.

\bigskip

\noindent
A successful precursor was an experiment 
by the MACHO Collaboration\cite{MACHO} in the 1990s.
In the 2020s, microlensing searches at the Vera Rubin
Observatory \cite{VRO} could repeat this success for the much
higher mass ranges of the MACHOs expected for the
dark matter inside the Milky Way.

\bigskip

\noindent
The MACHO collaboration, 1992-99, used the observatory at
Mount Stromlo near Canberra, Australia. it was a 1.27 m telescope
with two 16-Magapixel cameras. They showed that the technique
could be achieved successfully to discover MACHOs, as well as
confirming this prediction by Einstein's general relativity. The highest
duration of their more than a dozen light-curves was 230 days
corresponding to a mass close to $10 M_{\odot}$.

\bigskip

\noindent
An attempt was made to use the Blanco 4m telescope
at Cerro Tololo, Chile with the DECam having 570 Megapixels in
order to find light-curves with durations of two years or more,
and hence, by Eq.(\ref{duration}), lenses with $M > 100M_{\odot}$.
The longer durations led, however, to crowding in the field of view
such that it was impracticable to track a specific target star.

\bigskip

\noindent
A more powerful telescope under construction at Cerro Panchon,
also in Chile, is the Vera Rubin Observatory\cite{VRO} expected to start
taking data in 2023. Its telescope is 8.4 metres and its camera
has 3.2 Gigapixels, both significantly larger, and we can reasonably hope that it can microlens
multi-year-duration light curves and possibly confirm the existence of PIMBHs in the
Milky Way.

\section{Cosmic Infrared Background}

\noindent
At large red-shifts $Z > 15$,  a population of PBHs would be expected to accrete matter
and emit in X-ray and UV radiation which will be redshifted into the CIB
to be probed for the first time by the James Webb Space Telescope \cite{JWST}
which could therefore provide support for PBH formation.

\bigskip

\noindent
Analysis of a specific PBH formation model \cite{CHN} supports this idea that
the JWST observations in the infrared could provide relevant information
about whether PBHs really are formed in the early universe.

\bigskip

\noindent
This is important because although we have plenty of evidence for the existence
of black holes, whether any of them is primordial is not known.
The gravitational wave detectors\cite{LIGO} LIGO, VIRGO and KAGRA have
discovered mergers in black hole binaries with initial black holes in the mass
range $3 - 85 M_{\odot}$. We suspect that all or most of these are not
primordial but that is only conjecture.

\bigskip

\noindent
The supermassive black holes at galactic centres, including Sgr A* at the
centre of the Milky Way, are well established and are primordial in our
toy model. Whether that is the case in Nature is unknown.

\bigskip

\noindent
Because of the no-hair theorem that black holes are completely characterised
by their mass, spin and electric charge (usually taken to be zero), there is
no way to tell directly whether a given black hole is primordial or the result
of gravitational collapse of a star.

\bigskip

\noindent
The distinction between a primordial and a non-primordial black hole can be made
only from knowledge of its history. For example, if it existed before star formation,
it must be primordial. The infra-red data from JWST might be able to provide 
useful insight
into this central question.

\section{Discussion}

\noindent
It is familiar to study a mass-energy pie-chart of the universe with
approximately 5\% baryonic normal matter, 25\% dark matter and 70\%
dark energy. The entropy pie-chart is very different if the toy model
considered in this papers resembles Nature. The slices corresponding
to normal matter and dark energy are extremely thin and the pie
is essentially all dark matter.

\bigskip

\noindent
In this article we have attempted to justify better the discussion of our previous
2018 paper\cite{Frampton2018} which argued that entropy and the second law
applied to the early universe provide a {\it raison d'\^{e}tre} for the dark matter. 
In \cite{Frampton2015}
and \cite{ChaplineFrampton2016} we proposed that the dark matter constituents
in the Milky Way are PIMBHs.

\bigskip

\noindent
Here we have included the supermassive PSMBHs at the galactic centres as
a second tier of dark matter with a similar primordial origin to replace the conventional
wisdom that SMBHs arise from accretion and merging of black holes which arise
from gravity collapse of stars.

\bigskip

\noindent
We have gone one step further and discussed a third tier of the extremely
massive PEMBHs, more massive than galaxies, whose entropy far exceeds
that of the PIMBHs and PSMBHs. If this is correct then although normal matter
contributes as much as 5\% of the mass-energy pie-chart of the universe, its contribution
to an entropy pie-chart is truly infinitesimal.

\bigskip

\noindent
Since it has never been observed except by its gravity, it does seem
most likely that dark matter has no direct or even indirect connection to the standard model of
strong and electroweak interactions in particle theory, including extensions thereof
aimed to ameliorate problems with naturalness existing therein
with respect to the Higgs boson and the strong CP problem.

\bigskip

\noindent
The three clues we have mentioned in the Introduction, the dominance of black
holes in the entropy inventory, the CMB spectrum and the holographic entropy
maximum all hint toward PBHs as the dark matter constituent.

\bigskip

\noindent
One ambiguity is whether the maximum entropy limit suggested by
holography should be saturated in
which case the mass function for the PEMBHs must be extended to high values.

\section*{Acknowledgements}

\noindent
We thank the Department of Physics at the University
of Salento for an affiliation.  We thank G.F. Chapline and G. 't Hooft
for useful discussions.

\end{document}